\documentclass[prl,twocolumn]{revtex4}
\usepackage{amsfonts}
\usepackage{tabularx}
\usepackage{amsmath}
\usepackage{amssymb}
\usepackage{graphicx}
\usepackage{dcolumn}
\usepackage{bm}
\usepackage{ulem}

\begin{document}

\title{Topological Superfluids with Finite Momentum Pairing and Majorana
Fermions}
\author{Chunlei Qu$^{1}$}
\thanks{These authors contributed equally to this work}
\author{Zhen Zheng$^{2}$}
\thanks{These authors contributed equally to this work}
\author{Ming Gong$^{3}$}
\thanks{Email: skylark.gong@gmail.com}
\author{Yong Xu$^{1}$}
\author{Li Mao$^{1}$}
\author{Xubo Zou$^{2}$}
\thanks{Email: xbz@mail.ustc.edu.cn}
\author{Guangcan Guo$^{2}$}
\author{Chuanwei Zhang$^{1}$}
\thanks{Email: chuanwei.zhang@utdallas.edu}

\begin{abstract}
\textbf{Majorana fermions, quantum particles that are their own
anti-particles, are not only of fundamental importance in elementary
particle physics and dark matter, but also building blocks for
fault-tolerant quantum computation. Recently Majorana fermions have been
intensively studied in solid state and cold atomic systems. These studies
are generally based on superconducting pairing with zero total momentum. On
the other hand, finite total momentum Cooper pairings, known as
Fulde-Ferrell-Larkin-Ovchinnikov (FFLO) states, were widely studied in many
branches of physics. However, whether FFLO superconductors can support
Majorana fermions has not been explored. Here we show that Majorana fermions
can exist in certain types of gapped FFLO states, yielding a new quantum
matter: topological FFLO superfluids/superconductors. We demonstrate the
existence of such topological FFLO superfluids and the associated Majorana
fermions using spin-orbit coupled degenerate Fermi gases and derive their
parameter regions. The implementation of topological FFLO superconductors in
semiconductor/superconductor heterostructures are also discussed.}
\end{abstract}

\affiliation{$^{1}$Department of Physics, The University of Texas at Dallas, Richardson,
TX, 75080 USA \\
$^{2}$Key Laboratory of Quantum Information, University of Science and
Technology of China, Hefei, Anhui, 230026, People's Republic of China \\
$^{3}$Department of Physics, The Chinese University of Hong Kong, Shatin, N.T., Hong
Kong, China}
\maketitle

%\pacs{03.75.Ss, 74.20.Fg, 74.70.Tx, 03.67.Lx}

% Ming Gong verifed at 7.1.2013, please select several proper codes from the following candidates.
% 03.75.Ss Degenerate Fermi gases
% 74.20.Fg: BCS theory and its development
% 74.70.Tx: Heavy-fermion superconductors
% 74.25.Dw: Superconductivity phase diagrams
% 03.67.Lx Quantum computation architectures and implementations
% 03.65.Vf Phases: geometric; dynamic or topological
% 74.20.-z Theories and models of superconducting state

Topological superconductors and superfluids are exotic quantum matters that
host topological protected excitations, such as robust edge modes and
Majorana Fermions (MFs) with non-Abelian exchange statistics \cite{Wilczek}.
MFs are important not only because of their fundamental role in elementary
particle physics and dark matters \cite{Hisano}, but also their potential
applications in fault-tolerant topological quantum computation \cite{TQC}.
Recently some exotic systems, such as $\nu =5/2$ fractional quantum Hall
states \cite{TQC}, chiral \textit{p}-wave superconductors/superfluids \cite%
{TQC}, heterostructure composed of $s$-wave superconductors and
semiconductor nanowires (nanofilms) or topological insulators \cite{Fu,
JSau, Roman, Oreg, Alicea, Lee, Mao}, etc., have been proposed as systems
supporting MFs. Following the theoretical proposals, exciting experimental
progress for the observation of MFs has been made recently in semiconductor
\cite{Mourik, Deng, Das, Rokhinson} or topological insulator
heterostructures \cite{YCui}, although unambiguous experimental evidence for
MFs is still lacked.

These theoretical and experimental studies are based on the superconducting
Cooper pairing ($s$-wave or chiral $p$-wave) with zero total momentum, that
is, the pairing is between two fermions with opposite momenta $\mathbf{k}$
and $\mathbf{-k}$ (denoted as BCS pairing hereafter). On the other hand, the
superconducting pairing can also occur between fermions with finite total
momenta (pairing between $\mathbf{k}$ and $\mathbf{-k+Q}$) in the presence
of a Zeeman field, leading to spatially modulated superconducting order
parameters in real space, known as FFLO states. The FFLO states were first
predicted in 1960s \cite{FF64, LO64}, and now are a central concept for
understanding exotic phenomena in many different systems \cite%
{FFLOreview,FFLO2,FFLO3,FFLO1,Parish, HuPRA}. A natural question to ask is
whether MFs can also exist in a FFLO superconductor or superfluid?

In this Letter, we propose that FFLO superconductors/superfluids may support
MFs if they possess two crucial elements: gapped bulk quasi-particle
excitations and nontrivial Fermi surface topology. These new quantum states
are topological FFLO superconductors/superfluids. In this context,
traditional gapless FFLO states induced by a large Zeeman field do not fall
into this category. Here we propose a possible platform for the realization
of topological FFLO superfluids using two-dimensional (2D) or
one-dimensional (1D) spin-orbit (SO) coupled degenerate Fermi gases subject
to in-plane and out-of-plane Zeeman fields. Recently, the SO coupling and
Zeeman fields for cold atoms have already been realized in experiments \cite%
{Ian, Pan, Peter, Zhang, Martin}, which provide a completely new avenue for
studying topological superfluid physics. It is known that SO coupled
degenerate Fermi gases with an out-of-plane Zeeman field support MFs with
zero total momentum pairing \cite{CW,Jiang,Gong,Melo}. We find in suitable
parameter regions the in-plane Zeeman field can induce the finite total
momentum pairing \cite{Zheng, WYi, Hui, Lee2}, while still keeps the
superfluid gapped and preserves its Fermi surface topology. The region for
topological FFLO superfluids depends not only on the chemical potential,
pairing strength, but also on the SO coupling strength, total momentum and
effective mass of the Cooper pair, as well as the orientation and magnitude
of the Zeeman field, thus greatly increases the tunability in experiments.
Finally, the potential implementation of the proposal in
semiconductor/superconductor heterostructures is also discussed.

{\LARGE \textbf{Results}}

\begin{figure}[tbp]
\centering
\includegraphics[width=3.2in]{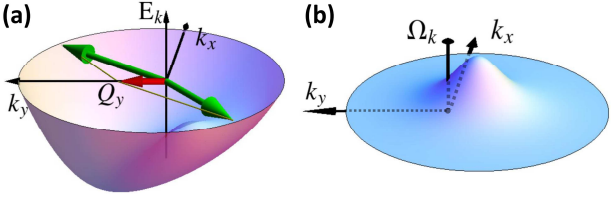}
\caption{\textbf{Single particle band structure and Berry curvature}. (a)
Energy  dispersion of the lower band. The green arrows represent the momenta
of a Cooper pair of two atoms on the asymmetric Fermi surface. The red arrow
represents the total finite momentum of the paring, which is along the
deformation direction of the Fermi surface. (b) Berry curvature of the lower
band, whose peak is shifted from the origin by $h_{x}/\protect\alpha $ along
the $k_{y}$ direction. }
\label{fig-FS}
\end{figure}

\textbf{System and Hamiltonian}: Consider a SO coupled Fermi gas in the $xy$
plane with the effective Hamiltonian
\begin{equation}
H=\sum_{\mathbf{k}\sigma \sigma ^{\prime }}c_{\mathbf{k},\sigma }^{\dagger
}H_{0}^{\sigma \sigma ^{\prime }}c_{\mathbf{k},\sigma ^{\prime }}+V_{\text{%
int}}  \label{eq-H}
\end{equation}%
where $H_{0}=\frac{\mathbf{k}^{2}}{2m}-\mu +\alpha \mathbf{k}\times \vec{%
\sigma}\cdot {\hat{e}_{z}}-\mathbf{h}\cdot \vec{\sigma}$, $\mathbf{k}%
=(k_{x},k_{y})$, $\alpha $ is the Rashba SO coupling strength, $\mathbf{h}%
=(h_{x},0,h_{z})$ is the Zeeman field and ${\vec{\sigma}}$ is the Pauli
matrices. $V_{\text{int}}=g{\sum }c_{\mathbf{k}_{1},\uparrow }^{\dagger }c_{%
\mathbf{k}_{2},\downarrow }^{\dagger }c_{\mathbf{k}_{3},\downarrow }c_{%
\mathbf{k}_{4},\uparrow }$ describes the $s$-wave scattering interaction,
where $g=-(\sum_{\mathbf{k}}\left( \mathbf{k}^{2}/m+E_{b}\right) ^{-1}{)}%
^{-1}$ is the scattering interaction strength, $E_{b}$ is the binding
energy, and $\mathbf{k}_{1}+\mathbf{k}_{2}=\mathbf{k}_{3}+\mathbf{k}_{4}$
due to the momentum conservation. Without in-plane Zeeman field $h_{x}$, the
Fermi surface is symmetric around $\mathbf{k}=0$, and the superfluid pairing
is between atoms with opposite momenta $\mathbf{k}$ and $-\mathbf{k}$. While
with both $h_{x}$ and SO coupling, the Fermi surface becomes asymmetric
along the $y$ direction (see Fig. \ref{fig-FS}a), and the pairing can occur
between atoms with momenta $\mathbf{k}$ and $-\mathbf{k}+\mathbf{Q}$. In
real space, such a finite total momentum pairing leads to a FF-type order
parameter $\Delta (\mathbf{x})=\Delta e^{i\mathbf{Q}\cdot \mathbf{x}}$,
where $\mathbf{Q}=(0,Q_{y})$ is parallel to the deformation direction of the
Fermi surface \cite{Zheng, WYi, Hui}. Notice that the energies of the
superfluids with total momentum $\mathbf{Q}$ and $\mathbf{-Q}$ are
nondegenerate, therefore FF phase with a single $\mathbf{Q}$, instead of LO
phase ($\Delta (\mathbf{x})=\Delta \cos (\mathbf{Q}\cdot \mathbf{x})$) where
pairing occurs at both $\pm \mathbf{Q}$, is considered here. Hereafter, if
not specified, FFLO superfluids refer to FF superfluids.

The dynamics of the system can be described by the following Bogliubov-de
Gennes (BdG) Hamiltonian in the mean-field level,
\begin{equation}
H_{\text{BdG}}(\mathbf{k})=%
\begin{pmatrix}
H_{0}({\frac{\mathbf{Q}}{2}}+\mathbf{k}) & \Delta \\
\Delta & -\sigma _{y}H_{0}^{\ast }({\frac{\mathbf{Q}}{2}}-\mathbf{k})\sigma
_{y}%
\end{pmatrix}%
,  \label{eq-bdg}
\end{equation}%
where the Nambu basis is chosen as $(c_{\mathbf{k}+\mathbf{Q}/2,\uparrow
},c_{\mathbf{k}+\mathbf{Q}/2,\downarrow },c_{-\mathbf{k}+\mathbf{Q}%
/2,\downarrow }^{\dagger },-c_{-\mathbf{k}+\mathbf{Q}/2,\uparrow }^{\dagger
})^{T}$. The gap, number and momentum equations are solved self-consistently
to obtain $\Delta $, $\mu $ and $\mathbf{Q}$, see Methods, through which we
determine different phases.

\textbf{Physical mechanism for topological FFLO phase}: Without SO coupling,
the orientation of the Zeeman field does not induce any different physics
due to SU(2) symmetry. The presence of both $h_{x}$ and SO coupling breaks
this SU(2) symmetry, leading to a Fermi surface without inversion symmetry,
see Fig. \ref{fig-FS}a. Here, $h_{x}$ deforms the Fermi surface, leading to
FFLO Cooper pairings; while $h_{z}$ opens a gap between the two SO bands,
making it possible for the chemical potential to cut a single Fermi surface
for the topological FFLO phase. The Berry curvature of the lower band reads
as
\begin{equation}
\Omega _{\mathbf{k}}={\frac{\alpha ^{2}h_{z}}{2(\alpha ^{2}k_{x}^{2}+(\alpha
k_{y}+h_{x})^{2}+h_{z}^{2})^{3/2}}}.
\end{equation}%
Note that $h_{x}$ shifts the peak of Berry curvature from $\mathbf{k}=0$ to $%
(0,-h_{x}/\alpha )$ (denoted by arrow in Fig. \ref{fig-FS}b). When atoms
scatter from $\mathbf{k}$ to $\mathbf{k}^{{\prime }}$ on the Fermi surface,
they pick up a Berry phase, whose accumulation around the Fermi surface $%
\theta =\int d^{2}\mathbf{k}\Omega _{\mathbf{k}}\approx \pi $. Such Berry
phase modifies the effective interaction from $s$-wave ($V_{\mathbf{k}%
\mathbf{k^{\prime }}}\sim g$ is a constant) to $s$-wave plus asymmetric $p$%
-wave
\begin{equation}
V_{\mathbf{k}\mathbf{k^{\prime }}}\sim g\left( ke^{-i\theta _{\mathbf{k}}}+{%
\frac{h_{x}}{\alpha }}\right) \left( k^{\prime }e^{i\theta _{\mathbf{k}%
^{\prime }}}-{\frac{h_{x}}{\alpha }}\right)
\end{equation}%
on the Fermi surface. Here we recover the well-known chiral $p_{x}+ip_{y}$
pairing \cite{CW} in the limit $h_{x}=0$. The in-plane Zeeman field here
creates an effective $s$-wave pairing component (although still hosts MFs),
and the effective pairing is reminiscent to the ($s$+$p$)-wave pairing in
some solid materials \cite{Yuan}.

\textbf{Parameter region for MFs}: The BdG Hamiltonian (\ref{eq-bdg})
satisfies the particle-hole symmetry $\Xi =\Lambda \mathcal{K}$, where $%
\Lambda =i\sigma _{y}\tau _{y}$, $\mathcal{K}$ is the complex conjugate
operator, and $\Xi ^{2}=1$. The parameter region for the MFs is determined
by the topological index $\mathcal{M}=\text{sign}(\text{Pf}\{\Gamma \})$,
where Pf is the Pfaffian of the skew matrix $\Gamma =H_{\text{BdG}%
}(0)\Lambda $. $\mathcal{M}=-1(+1)$ corresponds to the topologically
nontrivial (trivial) phase \cite{Parag}. The topological phase exists when
\begin{equation}
h_{z}^{2}+\bar{h}_{x}^{2}>\bar{\mu}^{2}+\Delta ^{2}\text{,}\quad \alpha
h_{z}\Delta \neq 0\text{,}\quad E_{g}>0,  \label{eq-parameter}
\end{equation}%
where $\bar{h}_{x}=h_{x}+\alpha Q_{y}/2$ and $\bar{\mu}=\mu -Q_{y}^{2}/8m$. $%
E_{g}=\text{min}(E_{\mathbf{k},s})$ defines the bulk quasi-particle
excitation gap of the system with $E_{\mathbf{k},s}$ as the particle
branches of the BdG Hamiltonian (\ref{eq-bdg}). The first condition reduces
to the well-known $h_{z}^{2}>\Delta ^{2}+\mu ^{2}$ in BCS topological
superfluids \cite{Mourik, Deng, Das, Rokhinson, Roman, Oreg, Gong}. The last
condition ensures the bulk quasi-particle excitations are gapped to protect
the zero energy MFs in the topological regime. The SO coupling and the FFLO
vector shift the effective in-plane Zeeman field and the chemical potential.
In contrast, in the BCS topological superfluids, the SO coupling strength,
although required, does not determine the topological boundaries. Our system
therefore provides more knobs for tuning the topological phase transition.
To further verify condition (\ref{eq-parameter}), we calculate the Chern
number in the hole branches $\mathcal{C}=$ $\sum_{n}\mathcal{C}_{n}$ in the
gapped superfluids \cite{Parag}, and confirm $\mathcal{C}=+1$ when Eq. (5)
is satisfied and $\mathcal{C}=0$ otherwise. Here $\mathcal{C}_{n}=\frac{1}{%
2\pi }\int d^{2}k\Gamma _{n}$ is the Chern number, $\Gamma _{n}=-2$Im$%
\left\langle \frac{\partial \Psi _{n}}{\partial k_{x}}|\frac{\partial \Psi
_{n}}{\partial k_{y}}\right\rangle $ is the Berry curvature \cite{Xiao}, and
$\left\vert \Psi _{n}\right\rangle $ is the eigenstate of two hole bands of
the BdG Hamiltonian (\ref{eq-bdg}).

\begin{figure}[tbp]
\centering
\includegraphics[width=3in]{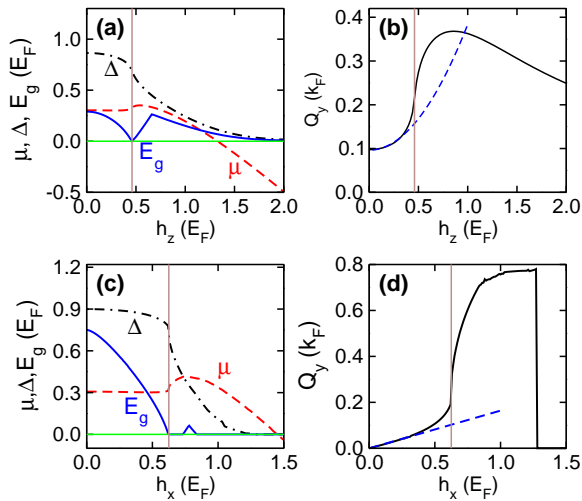}
\caption{\textbf{The order parameter $\Delta $, chemical potential $\protect%
\mu $, bulk quasi-particle gap $E_{g}$, and FFLO vector $Q_{y}$ as a
function of Zeeman fields}. In (b) and (d), the dashed lines are the best
fitting with quadratic and linear functions in the small Zeeman field
regime, respectively. In (a) and (b), $h_{x}=0.2E_{F}$, while in (c) - (d), $%
h_{Z}=0.2E_{F}$. Other parameters are $E_{b}=0.4E_{F}$, $\protect\alpha %
K_{F}=1.0E_{F}$. The vertical lines mark the points where the Pfaffian
changes the sign.}
\label{fig-Delta}
\end{figure}

The transition from non-topological to topological phases defined by Eq. (%
\ref{eq-parameter}) can be better understood by observing the close and
reopen of the excitation gap $E_{g}$, which is necessary to change the
topology of Fermi surface. In Fig. \ref{fig-Delta}, we plot the change of $%
E_{g} $ along with the order parameter $|\Delta |$, the chemical potential $%
\mu $, and the FF vector $\mathbf{Q}$ as a function of Zeeman fields. For a
fixed $h_{x}$ but increasing $h_{z}$, $E_{g}$ may first close and then
reopen (Fig. \ref{fig-Delta}a), signalling the transition from
non-topological to topological gapped FFLO superfluids ($Q_{y}$ is finite
for all $h_{z}$, see Fig. \ref{fig-Delta}b). For a fixed $h_{z}$, the
superfluid is gapped and $Q_{y}\propto h_{x}$ for a small $h_{x}$ (see Fig. %
\ref{fig-Delta}d), thus any small $h_{x}$ can transfer the gapped BCS
superfluids at $h_{x}=0$ to FFLO superfluids. However, such a small $h_{x}$
does not destroy the bulk gap of BCS superfluids (topological or
non-topological), making gapped topological FFLO superfluids possible when
the system is initially in topological BCS superfluids without $h_{x}$. With
increasing $h_{x}$ (Fig. \ref{fig-Delta}c), $E_{g}$ may first close but does
not reopen immediately, signalling the transition from gapped FFLO
superfluids to gapless FFLO superfluids. For a small $h_{z}=0.2E_{F}$,
further increasing $h_{x}$ to $\sim 0.78E_{F}$, $E_{g}$ reopens again (Fig. %
\ref{fig-Delta}c), signalling the transition from gapless FFLO to gapped
topological FFLO superfluids. In this regime, $Q_{y}\sim 0.6K_{F}$, which is
not small. For a strong enough Zeeman field, the pairing may be destroyed
and the system becomes a normal gas.

\begin{figure}[tbp]
\centering
\includegraphics[width=3in]{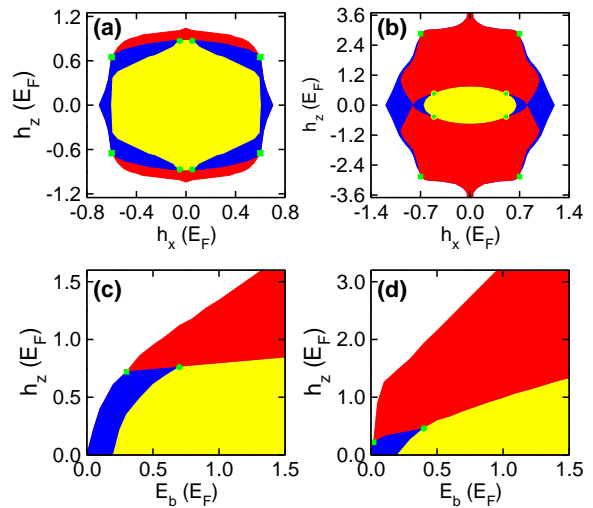}
\caption{(Color online). \textbf{Phase diagram of the FFLO superfluid}. The
phases are labelled with different colors: topological gapped FFLO
superfluid (red), non-topological gapped FFLO superfluid (yellow), gapless
FFLO superfluid (blue) and normal gas (white). Other parameters are: (a) $%
E_{b}=0.4E_{F}$, $\protect\alpha k_{F}=0.5E_{F}$; (b) $E_{b}=0.4E_{F}$, $%
\protect\alpha k_{F}=1.0E_{F}$; (c) $h_{x}=0.5E_{F}$, $\protect\alpha %
k_{F}=0.5E_{F}$; (d) $h_{x}=0.5E_{F}$, $\protect\alpha k_{F}=1.0E_{F}$. The
symbols in each panel are the tricritical points.}
\label{fig-Phases}
\end{figure}

The complete phase diagrams are presented in Fig. \ref{fig-Phases}. Since $%
Q_{y}$ and $h_{x}$ have the same sign, the phase diagram show perfect
symmetry in the $h_{x}-h_{z}$ plane. The BCS superfluids can only be
observed at $h_{x}=0$, hence are not depicted. With increasing SO coupling
strength, the topological FFLO phase is greatly enlarged through the
expansion to the normal gas phase. For a small SO coupling (Fig. \ref%
{fig-Phases}a), a finite $h_{z}$ is always required to create the
topological FFLO phase; In the intermediate regime (Fig. \ref{fig-Phases}b)
we find an interesting parameter regime where the topological FFLO phase can
be reached with an extremely small $h_{z}$ around $h_{x}\sim 0.8E_{F}$.
However, the topological FFLO phase can never be observed at $h_{z}=0$, as
analyzed before from the Berry curvature and Chern number. From Fig. \ref%
{fig-Phases}a-b we see that the topological gapped FFLO phase can be
mathematically regarded as an adiabatic deformation of the topological BCS
superfluids by an in-plane Zeeman field, although their physical meaning are
totally different. In Fig. \ref{fig-Phases}c-d, we see that the gapless FFLO
phase can be observed at small binding energy and small $h_{z}$, while for
large enough binding energy, the system can be either topological or
non-topological gapped phases. In this regime, $E_{g}\sim \sqrt{\mu
^{2}+\Delta ^{2}}-\sqrt{h_{x}^{2}+h_{z}^{2}}$, where $\mu \sim E_{F}-E_{b}/2$%
, and $\Delta ^{2}\sim 2E_{F}E_{b}$, thus $h_{z}\propto E_{b}$ is required
to close and reopen $E_{g}$ (see Fig. \ref{fig-Phases}c-d).

The tricritical points marked by symbols in Fig. \ref{fig-Phases} are
essential for understanding the basic structure of the phase diagram. Along
the $h_{z}$ axis, the system only supports gapped BCS superfluids
(topological or non-topological) and normal gas \cite{Gong}, while along the
$h_{x}$ axis the system only supports trivial FFLO superfluids and normal
gas \cite{Zheng, WYi, Hui}. So the adiabatic connection between the
topological BCS superfluids and trivial FFLO phases is impossible, and there
should be some points to separate different phases, which are exactly the
tricritical points. In our model the transition between different phases is
of first-order process. The existence of tricritical point here should be in
stark contrast to the tricritical point at finite temperature in the same
system without SO coupling, which arises from the accidental intersection of
first and second order transition lines \cite{Parish}. Therefore the
tricritical points in Fig. \ref{fig-Phases} cannot be removed, although
their specific positions vary with the system parameters.

\textbf{Chiral edge modes}: The topological FFLO superfluids support exotic
chiral edge modes. To see the basic features more clear, we consider the
same model in a square lattice with the following tight-binding Hamiltonian,
\begin{equation}
H_{\text{L}}=H_{0}+H_{\text{Z}}+H_{\text{so}}+V_{\text{int}},  \label{eq-TB}
\end{equation}%
where $H_{0}=-t\sum_{\langle i,j\rangle ,\sigma }c_{i\sigma }^{\dagger
}c_{j\sigma }-\mu \sum_{i\sigma }n_{i\sigma }$, $H_{\text{Z}%
}=-h_{x}\sum_{i}(c_{i\uparrow }^{\dagger }c_{i\downarrow }+c_{i\downarrow
}^{\dagger }c_{i\uparrow })-h_{z}\sum_{i}(n_{i\uparrow }-n_{i\downarrow })$,
$H_{\text{so}}=-\frac{\alpha }{2}\sum_{i}(c_{i-\hat{x}\downarrow }^{\dagger
}c_{i\uparrow }-c_{i+\hat{x}\downarrow }^{\dagger }c_{i\uparrow }+ic_{i-\hat{%
y}\downarrow }^{\dagger }c_{i\uparrow }-ic_{i+\hat{y}\downarrow }^{\dagger
}c_{i\uparrow }+\text{H.C})$, and $V_{\text{int}}=-U\sum_{i}n_{i\uparrow
}n_{i\downarrow }=\sum_{i}\Delta _{i}^{\ast }c_{i\downarrow }c_{i\uparrow
}+\Delta _{i}c_{i\uparrow }^{\dagger }c_{i\downarrow }^{\dagger }-|\Delta
_{i}|^{2}/U$, with $\Delta _{i}=-U\langle c_{i\downarrow }c_{i\uparrow
}\rangle $, $n_{i\sigma }=c_{i\sigma }^{\dagger }c_{i\sigma }$. Here $%
c_{i\sigma }$ denotes the annihilation operator of a fermionic atom with
spin $\sigma $ at site $i=(i_{x},i_{y})$. Hereafter, we use $t=1$ as the
basic energy unit. For more details, see Methods.

\begin{figure}[tbp]
\centering
\includegraphics[width=3in]{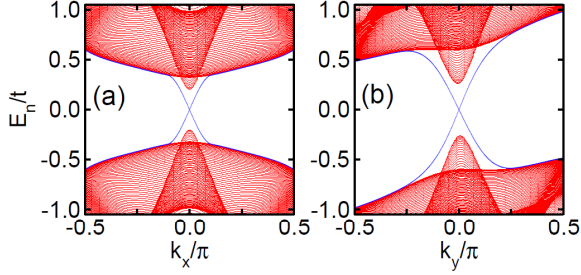}
\caption{\textbf{Chiral edge states of topological FFLO phases in a 2D strip}%
. The strip is along the $x$ direction (a); $y$ direction (b). The
parameters are $Q_{y}=-0.25$, $\protect\mu =-4t$, $\protect\alpha =2.0t$, $%
\Delta =1.0t$, $h_{z}=-1.2t$, $h_{x}=-0.3t$.}
\label{fig-edgestate}
\end{figure}

In the following, we only present the chiral edge states in the topological
gapped FFLO superfluid regime, and assume $\Delta _{i}=\Delta
e^{iQ_{y}i_{y}} $. We consider a 2D strip with width $W=200$, and the
results for the strip along $x$ and $y$ directions in the topological FFLO
phase are presented in Fig. \ref{fig-edgestate}. The linear dispersion of
the edge states reads as
\begin{equation}
H_{\text{edge}}=\sum_{k}v_{L}\psi _{kL}^{\dagger }k\psi _{kL}-v_{R}\psi
_{kR}^{\dagger }k\psi _{kR},
\end{equation}%
where $L$ and $R$ define the left and right edges of the strip, and $v_{L}$
and $v_{R}$ are the corresponding velocities. We have also confirmed that
the wavefunctions of the edge states are well localized at two edges. For a
strip along the $x$ direction, the particle-hole symmetry as well as the
discrete $\mathbb{Z}_{2}$ symmetry for $k_{x}\rightarrow -k_{x}$ ensure the
eigenenergies of Eq. \ref{eq-TB} always come in pairs ($E_{k}$,$-E_{k}$),
thus $v_{R}=v_{L}$. However, when the strip is along the $y$ direction
(parallel to the FFLO momentum $\mathbf{Q}$), the eigenenergies no longer
come in pairs, therefore $v_{R}\neq v_{L}$. The two chiral edge states with
totally different velocities and density of states represent the most
remarkable feature of our model. The Chern number $C=1$ in our lattice
model, thus only one pair of chiral edge states can be observed.
\begin{figure}[tbp]
\centering
\includegraphics[width=3in]{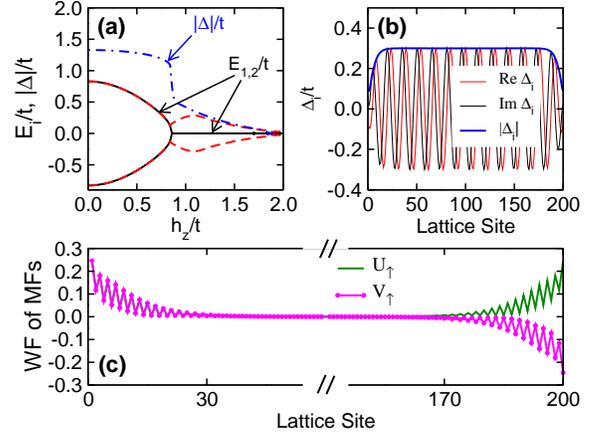}
\caption{\textbf{Majorana fermions in a 1D chain.} (a) The BdG
quasi-particle excitation energies ($E_{2}$, $E_{1}$, -$E_{1}$, -$E_{2}$)
and the order parameter; (b) The spatial profile of the FF type order
parameter obtained self-consistently. (c) The wavefunction (WF) of the
Majorana zero energy state $\left( U_{\uparrow },V_{\uparrow }\right) $ in
the 1D chain. $\left( U_{\downarrow },V_{\downarrow }\right) $ is similar
but with different amplitudes. The parameters are $\protect\alpha =2.0t$, $%
h_{x}=-0.5t$, $h_{z}=-1.2t$, $U=4.5t$, $\protect\mu =-2.25t$. }
\label{fig-mf}
\end{figure}

\textbf{MFs in 1D Chain}: Topological FFLO superfluid and associated MFs can
also be observed in 1D SO coupled Fermi gas when the Hamiltonian (\ref{eq-TB}%
) is restricted to 1D chain. In this case, the system is characterized by a $%
\mathbb{Z}_{2}$ invariant, which can be determined using the similar
procedure as discussed above. The only difference is that now not only $k=0$%
, but also $k=\pi $ needs be taken into account (see Methods). In Fig. \ref%
{fig-mf}a, we see Majorana zero-energy state protected by a large gap ($\sim
0.3t$) emerges in a suitable parameter region. The superfluid order
parameter (Fig. \ref{fig-mf}b) has the FF form. The local Bogoliubov
quasi-particle operator $\gamma (E_{n})=\sum_{i\sigma }u_{i\sigma
}^{n}c_{i\sigma }+v_{i\sigma }^{n}c_{i\sigma }^{\dagger }$, where the zero
energy wavefunction $\left( u_{i\uparrow }^{0},u_{i\downarrow
}^{0},v_{i\uparrow }^{0},v_{i\downarrow }^{0}\right) =\left( U_{i\uparrow
}e^{i\phi _{i\uparrow }},U_{i\downarrow }e^{i\phi _{i\downarrow
}},V_{i\uparrow }e^{-i\phi _{i\uparrow }},V_{i\downarrow }e^{-i\phi
_{i\downarrow }}\right) $ satisfies $u_{i\sigma }^{0}=v_{i\sigma }^{0\ast }$
at the left edge and $u_{i\sigma }^{0}=-v_{i\sigma }^{0\ast }$ at the right
edge (see Fig. \ref{fig-mf}c). This state supports two local MFs at two
edges, respectively \cite{Roman}.

{\LARGE \textbf{Discussions}}

Our proposed topological FFLO phase may also be realized using
semiconductor/superconductor heterostructures. Recently, topological BCS
superconductors and the associated MFs have been proposed in such
heterostructures \cite{JSau, Roman, Oreg, Alicea} and some preliminary
experimental signatures have been observed \cite{Mourik, Deng, Das,
Rokhinson}. To realize a topological FFLO superconductor, the semiconductor
should be in proximity contact with a FFLO superconductor, which introduces
finite momentum Cooper pairs. The topological parameter region defined in
Eq. (5) still applies except that the order parameter, chemical potential
and FFLO vector are external independent parameters. The flexibility of Eq.
(5) makes it easier for tuning to the topological region with MFs. Because
the FFLO state can sustain in the presence of a large magnetic field, it
opens the possibility for the use of many semiconductor nanowires with large
spin-orbit coupling but small $g$-factors (e.g, GaSb, hole-doped InSb, etc.).

In summary, we propose that topological FFLO superfluids or superconductors
with finite momentum pairings can be realized using SO coupled $s$-wave
superfluids subject to Zeeman fields and they support exotic quasi-particle
excitations such as chiral edge modes and MFs. The phase transition to the
topological phases depends strongly on all physical quantities, including SO
coupling, chemical potential, Zeeman field and its orientations, paring
strength, FFLO vector $\mathbf{Q}$ and the effective mass of Cooper pairs
explicitly, which are very different from topological BCS
superfluids/superconductors that are intensively studied recently. These new
features not only provide more knobs for tuning topological phase
transitions, but also greatly enrich our understanding of topological
quantum matters. The topological FFLO phases have not been discussed before,
and the phases unveiled in this Letter represent a totally new quantum
matter.

{\LARGE \textbf{Methods}}

\textbf{Momentum space BdG equations}: The partition function at finite
temperature $T$ is $Z=\int \mathcal{D}[\psi ,\psi ^{\dagger }]e^{-S[\psi
,\psi ^{\dagger }]}$, where $S[\psi ,\psi ^{\dagger }]=\int d\tau d\mathbf{r}%
\sum_{\sigma =\uparrow ,\downarrow }\psi _{\sigma }(\mathbf{x})^{\dagger
}\partial _{\tau }\psi _{\sigma }(\mathbf{x})+H$, with $H$ defined in Eq. %
\ref{eq-H}, and $V_{\text{int}}=g\psi _{\uparrow }^{\dagger }\psi
_{\downarrow }^{\dagger }\psi _{\downarrow }\psi _{\uparrow }$ in real
space. The FFLO phase is defined as $g\langle \psi _{\downarrow }(\mathbf{x}%
)\psi _{\uparrow }(\mathbf{x})\rangle =\Delta e^{i\mathbf{Q}\cdot \mathbf{x}%
} $, where $\mathbf{Q}$ is the total momentum of the Cooper pairs and $%
\Delta $ is a spatially independent constant. Here the position dependent
phase of $\Delta (\mathbf{x})$ can be gauged out by the transformation $\psi
_{\sigma }\rightarrow \psi _{\sigma }e^{i\mathbf{Q}\cdot \mathbf{x}/2}$.
Integrating out the fermion field $\psi $ and $\psi ^{\dagger }$, we obtain $%
Z=\int \mathcal{D}\Delta e^{-S_{\text{eff}}}$, with effective action $S_{%
\text{eff}}=\int d{\tau }d\mathbf{r}{\frac{|\Delta |^{2}}{g}}-{\frac{1}{%
2\beta }}\ln \text{Det}\beta G^{-1}+\text{Tr}(H)$, where $\beta =1/T$, and $%
G^{-1}=\partial _{\tau }+H_{\text{BdG}}$. The order parameter, chemical
potential and FFLO vector $\mathbf{Q}$ are determined self-consistently by
solving the following equation set
\begin{equation}
{\frac{\partial S_{\text{eff}}}{\partial \Delta }}=0,\quad {\frac{\partial
S_{\text{eff}}}{\partial \mu }}=-\beta n,\quad {\frac{\partial S_{\text{eff}}%
}{\partial \mathbf{Q}}}=0.
\end{equation}%
In our model the deformation of Fermi surface is along the $y$ direction,
thus we have $\mathbf{Q}=(0,Q_{y})$, and only three parameters need be
determined self-consistently. We determine the different quantum phases
using the following criterion. When $E_{g}>0$, $\Delta \neq 0$, we have
gapped FFLO phases ($\mathcal{M}=-1$ ($\mathcal{C}=+1$) for topological, and
$M=+1$ ($C=0$) for non-topological). When there is a nodal line with $%
E_{g}=0 $ and $\Delta \neq 0$, we have gapless FFLO phases. When $\Delta =0$
(then $\mathbf{Q}=0$ is enforced), we get normal gas phases. It is still
possible to observe gapless excitations in the gapless FFLO phase regime,
however, we do not distinguish this special condition because gapless
excitations are not protected by gaps. In our numerics, the energy and
momentum are scaled by Fermi energy $E_{F}$ and its corresponding momentum $%
K_{F}$ in the case without SO coupling and Zeeman fields. The results in
Fig. \ref{fig-Delta} and Fig. \ref{fig-Phases} are determined at $%
n=K_{F}^{2}/2\pi $ and $T=0$.

\textbf{Real space BdG equations}: In the tight-binding model of (\ref{eq-TB}%
), the many-body interaction is decoupled in the mean-field approximation.
The particle number $n_{i\sigma }=c_{i\sigma }^{\dagger }c_{i\sigma }$ and
superfluid pairing $\Delta _{i}=-U\langle c_{i\downarrow }c_{i\uparrow
}\rangle $ are determined self-consistently for a fixed chemical potential.
Using the Bogoliubov transformation, we obtain the BdG equation
\begin{equation}
\sum_{j}%
\begin{pmatrix}
H_{ij\uparrow } & \alpha _{ij} & 0 & \Delta _{ij} \\
-\alpha _{ij} & H_{ij\downarrow } & -\Delta _{ij} & 0 \\
0 & -\Delta _{ij}^{\ast } & -H_{ij\uparrow } & -\alpha _{ij} \\
\Delta _{ij}^{\ast } & 0 & \alpha _{ij} & -H_{ij\downarrow }%
\end{pmatrix}%
\begin{pmatrix}
u_{j\uparrow }^{n} \\
u_{j\downarrow }^{n} \\
-v_{j\uparrow }^{n} \\
v_{j\downarrow }^{n}%
\end{pmatrix}%
=E_{n}%
\begin{pmatrix}
u_{j\uparrow }^{n} \\
u_{j\downarrow }^{n} \\
-v_{j\uparrow }^{n} \\
v_{j\downarrow }^{n}%
\end{pmatrix}%
,  \label{BdG}
\end{equation}%
where $H_{ij\uparrow }=-t\delta _{i\pm 1,j}-(\mu +h_z)\delta _{ij}$, $%
H_{ij\downarrow }=-t\delta _{i\pm 1,j}-(\mu -h_z)\delta _{ij}$, $\alpha
_{ij}=\frac{1}{2}(j-i)\alpha \delta _{i\pm 1,j}-h_x \delta_{i,j}$, $%
\left\langle \hat{n}_{i\sigma }\right\rangle =\sum_{n=1}^{2N}[|u_{i\sigma
}|^{2}f(E_{n})+|v_{i\sigma }|^{2}f(-E_{n})]$, $\Delta _{ij}=-U\delta
_{ij}\sum_{n=1}^{2N}[u_{i\uparrow }^{n}v_{i\downarrow }^{n\ast
}f(E_{n})-u_{i\downarrow }^{n}v_{i\uparrow }^{n\ast }f(-E_{n})]$, with $%
f(E)=1/\left( 1+e^{E/T}\right) $. In the tight-binding model, FF phase and
LO phase can be determined naturally, which depend crucially on the
parameters of the system as well as the position of the chemical potential.
The results in Fig. \ref{fig-edgestate} and Fig. \ref{fig-mf} are obtained
at $T=0$.

\textbf{Topological boundaries in lattice models:} To determine the
topological phase transition conditions, we transform the tight-binding
Hamiltonian to the momentum space in Eq. \ref{eq-bdg}. Here $\xi _{\mathbf{k}%
}$ is replaced by $-2t\cos (k_{x})-2t\cos (k_{y})-\mu $ for the kinetic
energy, and $k_{\alpha }$ by $\sin (k_{\alpha })$ for the SO coupling, where
$\alpha =x,y$. The topological boundary conditions can still be determined
by the Pfaffian of $\Gamma (\mathbf{K})=H_{\text{BdG}}(\mathbf{K})\Lambda $
at four nonequivalent points, $K_{1}=(0,0)$, $K_{2}=(0,\pi )$, $K_{3}=(\pi
,0)$, $K_{4}=(\pi ,\pi )$ when the system is gapped. At these special
points, $\Gamma (\mathbf{k})$ is a skew matrix. The topological phase is
determined by $\mathcal{M}=\prod_{i=1}^{4}\text{sign}(\text{Pf}(\Gamma
(K_{i})))=-1$. For uniform BCS superfluids, the Pfaffian at $K_{2}$ and $%
K_{3}$ are identical, thus only $K_{1}$ and $K_{4}$ are essential to
determine the topological boundaries. However, in our system, all four
points affect the topological boundaries, and the exact expression of $%
\mathcal{M}$ is too complex to present here. In 1D chain, there are only two
nonequivalent points at $K_{1}=0$ and $K_{2}=\pi $. We find $\text{Pf}%
(\Gamma (K_{1}))=\Delta ^{2}-(h_{z}-\mu -2t\cos (Q_{y}/2))(h_{z}+\mu +2t\cos
(Q_{y}/2))-(h_{x}+\alpha \sin (Q_{y}/2))^{2}$ , and $\text{Pf}(\Gamma
(K_{2}))=\Delta ^{2}-(h_{z}-\mu +2t\cos (Q_{y}/2))(h_{z}+\mu -2t\cos
(Q_{y}/2))-(h_{x}-\alpha \sin (Q_{y}/2))^{2}$. The topological index in the
gapped regime is determined by $\text{sign}(\text{Pf}(\Gamma (K_{1})))\text{%
sign}(\text{Pf}(\Gamma (K_{2})))$.

\textbf{Acknowledgement}

C.Q, Y.X, L.M, C.Z are supported by ARO (W911NF-12-1-0334), AFOSR
(FA9550-13-1-0045), and NSF-PHY (1104546). Z.Z., X.Z., and G.G. are
supported by the National 973 Fundamental Research Program (Grant No.
2011cba00200), the National Natural Science Foundation of China (Grant No.
11074244 and No. 11274295). M.G is supported in part by Hong Kong RGC/GRF
Project 401512, the Hong Kong Scholars Program (Grant No. XJ2011027) and the
Hong Kong GRF Project 401113.

\textbf{Author contributions} All authors designed and performed the
research and wrote the manuscript.

\textbf{Competing financial interests}

The authors declare no competing financial interests.

\end{document}